\documentclass[twocolumn, prd, showpacs, amsmath, amssymb]{revtex4-2}
\usepackage{amsmath, amssymb}
\usepackage{physics}
\usepackage{tikz}
\usepackage{enumitem}
\usepackage{graphicx}

\begin{document}

%\begin{frontmatter}
\title{Quantized Dissipation from the Inverse-Square Anomaly in a Non-Hermitian Klein--Gordon Field}
\author{Mansour Haghighat\footnote{e-mail:{m.haghighat@shirazu.ac.ir}}}
\author{Ali Nouri}
\affiliation{Physics Department, College of Sciences, Shiraz University, 71454, Shiraz, Iran}

%\author{M. Haghighat\corref{cor1}}
%\ead{m.haghighat@shirazu.ac.ir}
%\ead{mhaghighat44@gmail.com}
%\cortext[cor1]{Corresponding author.}
%\author{A. Nouri}
%\address{Physics Department, College of Sciences, Shiraz University, 71454, Shiraz, Iran}
\date{\today}

\begin{abstract}
We construct an exactly solvable relativistic model that embeds the anomalous inverse-square interaction into a non-Hermitian Klein--Gordon field theory through a purely imaginary, scale-invariant scalar potential. The stationary field equation reduces to an inverse-square Schrödinger-type problem with a quadratic spectral parameter. By imposing a perfectly absorbing boundary condition at the singularity, the Hermitian inverse-square anomaly is naturally continued to an open dissipative system, where the geometric tower of bound states is reinterpreted as a hierarchy of resonant decay modes. We present an explicit analytical derivation showing that the resonance decay rates obey a universal log-periodic scaling relation determined solely by the anomalous scaling exponent and robust against generic subleading perturbations. The construction establishes a direct connection between self-adjoint boundary conditions and their non-Hermitian absorbing counterparts through analytic continuation of the boundary phase, demonstrating that the underlying discrete scale invariance survives in the presence of irreversible absorption. The resulting framework provides a minimal, exactly solvable laboratory for studying quantum scale anomalies, boundary-condition-induced non-Hermiticity, and quantized dissipation in relativistic open quantum systems.

\end{abstract}
%\end{frontmatter}
\maketitle
\section{Introduction}
The inverse-square potential provides a paradigmatic example of a quantum anomaly. Although the Hamiltonian $H=-\nabla^2-\lambda/r^2$ is classically scale invariant, for sufficiently strong coupling the operator ceases to be essentially self-adjoint and develops the fall-to-the-center instability \cite{Case1950, deAlfaro1976}. Restoring self-adjointness requires the introduction of a dimensional parameter through boundary conditions, breaking scale invariance by dimensional transmutation. Despite its long history, the physical meaning of this parameter remains largely formal within Hermitian quantum mechanics.

At the same time, non-Hermitian quantum mechanics has established that effective non-unitary dynamics naturally describes open systems coupled to unobserved degrees of freedom, while global unitarity may still be recovered in an enlarged Hilbert space \cite{El-Ganainy2018}. This raises the possibility that certain quantum anomalies admit physically meaningful resolutions when embedded into dissipative relativistic frameworks.

Here we construct an exactly solvable non-Hermitian Klein–Gordon model that realizes the inverse-square anomaly as irreversible absorption at a singular boundary. Imposing a physically motivated ingoing boundary condition converts the fall-to-the-center instability into a discrete spectrum of complex energies with a universal log-periodic decay ladder. The spectrum defines an emergent Hawking-like effective temperature, establishing a direct and fully analytic connection between scale anomaly, non-Hermiticity, and quantized dissipation in flat spacetime.

Crucially, the geometric spacing of decay rates is fixed solely by the anomalous scaling exponent, rendering the dissipative spectrum universal and insensitive to microscopic regularization details or large-scale potential modifications. Owing to its analytic solvability, universality, and compatibility with engineered dissipative platforms, the model provides a minimal framework for exploring anomaly-driven dissipation in flat spacetime.

\section{Non-Hermitian Klein--Gordon embedding}
We consider a complex scalar field $\psi(t,\mathbf{r})$ governed by the Klein--Gordon equation in flat spacetime in the presence of a scalar potential $V(r)$,
\begin{equation}
\left[ \partial_t^2 - \nabla^2 + \big(m + V(r)\big)^2 \right]\psi(t,\mathbf{r}) = 0 ,
\label{KG_general}
\end{equation}
where $m$ is the bare mass and $V(r)$ is allowed to be complex. Scalar potentials enter as shifts of the mass term~\cite{2003Paper}, in contrast to vector potentials which couple through minimal substitution. When $V(r)$ is complex, the resulting field theory is explicitly non-Hermitian and describes an open quantum system with gain or loss.

We choose the scalar potential in the scale-invariant form
\begin{equation}
m + V(r) = \frac{i\gamma}{r}, \qquad \gamma \in \mathbb{R},
\label{imag_mass}
\end{equation}
or equivalently $V(r) = -m + i\gamma/r$. This choice exactly cancels the rest mass and replaces it by a purely imaginary, scale-invariant effective mass that diverges at the origin. While this exact cancellation may appear fine-tuned, we demonstrate in Sec.~\ref{sec:robustness} and \ref{app:absorber} that the resulting anomalous spectrum is robust against generic physical perturbations, and such effective mass tuning is standard in engineered synthetic systems. Physically, the singularity at $r=0$ acts as a perfect absorber: probability amplitude reaching the origin is irreversibly lost from the reduced field sector. The system thus provides a minimal realization of an open relativistic quantum theory with exact scale invariance.

For stationary configurations,
\begin{equation}
\psi(t,\mathbf{r}) = e^{-iEt}\phi(\mathbf{r}),
\end{equation}
Eq.~(\ref{KG_general}) reduces to the time-independent equation
\begin{equation}
\left[ -\nabla^2 - \frac{\gamma^2}{r^2} \right]\phi(\mathbf{r}) = E^2\phi(\mathbf{r}) .
\label{Sch_like}
\end{equation}
This equation has the form of a Schrödinger-type eigenvalue problem with an attractive inverse-square potential, with the spectral parameter given by $E^2$ rather than $E$. Importantly, Eq.~(\ref{Sch_like}) follows exactly from the relativistic Klein--Gordon theory and does not rely on any nonrelativistic approximation.

Equation~(\ref{Sch_like}) is manifestly scale invariant: under the transformation $r \to \lambda r$ and $E \to E/\lambda$, the equation remains unchanged. This continuous scale symmetry is the relativistic counterpart of the familiar conformal symmetry of the nonrelativistic inverse-square Hamiltonian. As in the standard anomalous problem, the attractive $1/r^2$ interaction leads to a singular radial operator that is not essentially self-adjoint for sufficiently large coupling. In the present setting, however, the interpretation is fundamentally different: the theory is explicitly non-Hermitian, and the singularity at the origin is endowed with a clear physical meaning as a one-way sink for probability.

It is crucial to emphasize that the appearance of the inverse-square interaction in Eq.~(\ref{Sch_like}) is not imposed by hand at the level of a nonrelativistic Hamiltonian, but emerges directly from a relativistic field equation with a complex scalar potential. This embedding allows the familiar inverse-square anomaly to be reinterpreted as a problem of boundary conditions and information flow in an open relativistic system.

\section{Self-adjoint extension, outgoing boundary condition, and quantized decay}
We now analyze the spectral problem associated with Eq.~(\ref{Sch_like}),
focusing on the behavior near the singular point $r=0$ and the role of boundary conditions in determining the physical spectrum.

\subsection{Radial reduction and singular behavior}
Separating variables in spherical coordinates,
\begin{equation}
\phi(\mathbf{r}) = \frac{u_\ell(r)}{r}Y_{\ell m}(\Omega),
\end{equation}
the radial function satisfies
\begin{equation}
\left[ -\frac{d^2}{dr^2} + \frac{\ell(\ell+1)-\gamma^2}{r^2} \right]u_\ell(r) = E^2 u_\ell(r).
\label{radial}
\end{equation}
Defining the effective attractive coupling
\begin{equation}
\alpha_\ell \equiv \gamma^2 - \ell(\ell+1),
\end{equation}
the radial operator 
\begin{equation}
H_\ell=-\frac{d^2}{dr^2}-\frac{\alpha_\ell}{r^2},
\label{radial-operator}
\end{equation}
ceases to be essentially self-adjoint on $C_0^\infty(0,\infty)$ when
\begin{equation}
\alpha_\ell > \frac{1}{4}.
\end{equation}
In this regime the Hermitian inverse-square problem admits a one-parameter family of self-adjoint extensions, reflecting an intrinsic ambiguity in the boundary condition at the singularity.

Near the origin, as is shown in \ref{app:anomaly}, the two independent solutions behave as
\begin{equation}
u_\ell(r) \sim r^{\frac{1}{2}\pm i\sigma_\ell},
\qquad
\sigma_\ell = \sqrt{\alpha_\ell - \frac{1}{4}} .
\label{near0}
\end{equation}
Both solutions are square integrable at $r=0$, making the specification of a boundary condition at the singularity unavoidable.
Note that this behavior differs fundamentally from the standard free-particle limit $u_\ell(r) \sim r^{\ell+1}$, because the strong attractive coupling $\gamma^2 > \ell(\ell+1) + 1/4$ ensures that the inverse-square potential dominates over the centrifugal barrier in the UV, shifting the indicial exponents from real integers to complex conjugates, see \ref{app:anomaly}.
\subsection{Outgoing boundary condition and absorbing singularity}
The Hamiltonian in the Hermitian inverse-square problem admits a one-parameter family of self-adjoint extensions.  Usually, the relative phase between the two solutions in Eq.~(\ref{near0}) is fixed by an arbitrary and real self-adjoint extension parameter. In the present non-Hermitian Klein--Gordon embedding, however, we would like to study 
a different physical realization. In fact, instead of a perfectly reflecting singularity, the singularity is modeled as
a perfectly absorbing boundary. Physically, such boundary conditions are commonly employed in effective descriptions of open systems, where the omitted microscopic degrees of freedom are represented by irreversible probability loss through the boundary.  
Mathematically, as demonstrated in \ref{app:absorber}, this boundary condition can be viewed as the natural non-Hermitian continuation of Case's self-adjoint extension. By analytically continuing Case's real boundary phase $\delta$ to the complex plane ($\delta = \delta_R + i\delta_I$), the imaginary part $\delta_I$ acts as a continuous dial for the absorption strength. The perfectly absorbing boundary condition corresponds to the maximally dissipative limit ($\delta_I \to +\infty$) of this one-parameter family, which corresponds to the limiting case of complete absorption, appropriate when the effective coupling to unresolved short-distance degrees of freedom is sufficiently strong.

The general solution near the origin is a superposition $u_\ell(r) = C_- r^{\frac{1}{2}-i\sigma_\ell} + C_+ r^{\frac{1}{2}+i\sigma_\ell}$. The radial probability current for this superposition is strictly given by $j_r = \sigma_\ell \left( |C_+|^2 - |C_-|^2 \right)$. For the singularity to act as a perfect, irreversible absorber, there must be no reflected outward flux. This physical requirement dictates that the outward-traveling component must vanish entirely, imposing the boundary condition $C_+ = 0$. 

Consequently, the near-origin solution is uniquely selected as:
\begin{equation}
u_\ell(r) \propto r^{\frac{1}{2}-i\sigma_\ell},
\end{equation}
which corresponds to a strictly inward-directed probability flux ($j_r = -\sigma_\ell |C_-|^2 < 0$) as $r\to0$. In this way, the self-adjoint extension ambiguity of the Hermitian formulation is bypassed, and the boundary condition is fixed in a physically unique manner by the requirement of pure absorption.
\subsection{Log-periodic quantization of complex energies}
At large $r$, solutions of Eq.~(\ref{radial}) behave as superpositions of oscillatory waves. Matching the purely ingoing solution at the singularity to the asymptotic large-$r$ behavior yields a discrete set of complex resonant energies,
\begin{equation}
E_n = E_0 \exp\left(-\frac{\pi n}{\sigma_\ell}\right),
\qquad n\in\mathbb{Z},
\label{En}
\end{equation}
where $E_0$ is a complex reference scale fixed by short-distance matching at a regularization scale $r=r_0$. The spectrum therefore forms a \emph{log-periodic tower}, characteristic of quantum systems with continuous scale invariance broken by an irreversible boundary condition. The explicit analytical derivation of this quantization condition and its inherent robustness is detailed in \ref{app:absorber}.

Writing
\begin{equation}
E_n = \omega_n - \frac{i}{2}\Gamma_n ,
\end{equation}
the decay widths satisfy
\begin{equation}
\Gamma_{n+1} = \Gamma_n \exp\left(-\frac{\pi}{\sigma_\ell}\right),
\label{Gamma_ladder}
\end{equation}
demonstrating that dissipation proceeds in a \emph{quantized geometric progression} rather than continuously.  Consequently, the resonance widths inherit the discrete scale invariance of the anomalous inverse-square potential.

\subsection{Selection of the physical energy branch}
\label{sec:branch}
Since Eq.~(\ref{radial}) determines the spectral parameter $E^2$, the resonance energy is obtained by choosing one of the two square-root branches of the complex eigenvalue. In a non-Hermitian framework, this corresponds to choosing between exponentially decaying ($\mathrm{Im}\,E < 0$) and exponentially growing ($\mathrm{Im}\,E > 0$) time evolution. We explicitly select the decaying branch based on the physical requirement of asymptotic stability and causality. The growing branch represents a runaway solution that is unphysical for a stationary open system modeled as a pure absorber; such a branch would only become relevant if non-linear saturation effects (beyond the scope of this linear effective theory) were included to stabilize the growth. Thus, the physical spectrum is strictly restricted to $\mathrm{Im}\,E_n < 0$.

\section{Robustness and Physical Realizations}
\label{sec:robustness}

\subsection{Robustness against perturbations}
A natural concern regarding the exact cancellation of the bare mass term in Eq.~(\ref{imag_mass}) is whether this represents an artificial fine-tuning. We address this from both physical and theoretical perspectives. 

Physically, in engineered quantum systems (e.g., photonic lattices or ultracold atoms), the ``mass'' term is an effective parameter determined by the band structure or dispersion relation, not a fundamental constant. Tuning this effective parameter to achieve specific resonance or cancellation conditions is a standard, realistic experimental capability. Furthermore, even if practical limitations result in a small residual deviation $\delta m$ from perfect cancellation, this perturbation is subleading compared to the divergent $i\gamma/r$ term as $r \to 0$. Consequently, such deviations cannot affect the UV properties of the system: the dominant $1/r^2$ singularity, the anomalous scaling exponent $\sigma_\ell$, and the resulting geometric decay ratio remain strictly unchanged. The exact cancellation is thus not a fragile fine-tuning but rather defines a stable universality class for the infrared behavior, while the UV anomaly is robust against any regular perturbation.
To confirm this stability rigorously, note that any perturbation satisfying $\delta V(r) \ll \gamma/r$ as $r \to 0$ is subleading compared to the dominant $1/r^2$ singularity. Standard theorems on singular differential operators (Weyl’s limit point/limit circle classification~\cite{ReedSimon}) guarantee that such perturbations do not alter the deficiency indices or the leading indicial exponent $\sigma_\ell$. They merely shift the short-distance matching phase, thereby modifying the overall reference scale $E_0$ while leaving the geometric ratio $\exp(-\pi/\sigma_\ell)$ strictly invariant. A detailed discussion is provided in \ref{app:absorber}.

\subsection{Effective Description and Possible Physical Realizations}
The present model should be viewed as an exactly solvable effective description of an open relativistic wave system rather than as a fundamental field theory with a microscopic imaginary scalar interaction. The purely imaginary 1/r potential is introduced as an effective representation of irreversible absorption after eliminating unresolved environmental degrees of freedom. Its main purpose is to isolate, in an analytically tractable setting, the universal consequences of the inverse-square quantum anomaly and absorbing boundary conditions. Once interpreted in this effective sense, similar mathematical structures may be realized in a variety of engineered wave system, some representative examples of which are discussed below.

\textbf{1. Paraxial Optics and Photonic Lattices:} The paraxial wave equation for light propagation in a medium with a complex refractive index $n(r) = n_0 + i n_I(r)$ is mathematically isomorphic to the Schrödinger/Klein-Gordon equation. An imaginary potential $V(r) \propto i/r$ directly maps to a radially graded gain/loss profile $n_I(r) \propto 1/r$. Such profiles can be engineered in optical fibers or waveguides via specific doping or pumping profiles \cite{El-Ganainy2018, Chang2018}. In this context, the singularity at $r=0$ represents a highly localized, engineered loss channel, and the log-periodic spectrum would manifest as discrete resonance rings in the transmission spectrum.

\textbf{2. Effective Open Quantum Systems:} In the framework of effective field theory, an imaginary potential arises naturally when a subsystem is coupled to an unobserved continuum of states (e.g., via the Feshbach projection formalism). The $1/r$ profile can emerge from the density of states of the environment in specific geometries, providing a rigorous effective description of irreversible absorption without requiring fundamental non-Hermiticity.  Within this interpretation, the imaginary $1/r$ potential represents an effective coarse-grained description of irreversible probability loss into unresolved short-distance degrees of freedom, consistent with the absorbing boundary condition imposed in the present work.

\section{Emergent Hawking-like temperature}
The logarithmic decay spectrum,
\begin{equation}
\ln\Gamma_n=\ln(2|E_0|)-\frac{\pi}{\sigma_\ell}n,
\end{equation}
defines a single emergent energy scale. By direct analogy with black-hole quasi-normal spectra, we define the \emph{effective temperature}
\begin{equation}
T_{\rm eff}\equiv\frac{1}{2\pi k_B}\frac{|E_0|}{\sigma_\ell}\sim
\frac{1}{2\pi k_B}\frac{1}{\sigma_\ell r_0}.
\end{equation}
This temperature is purely kinematic, arising from scale invariance and one-way boundary conditions, without gravitational dynamics.

At a more fundamental level, the results demonstrate that the inverse-square anomaly controls not only bound-state instabilities but also the universal structure of dissipative spectra in open relativistic systems. The geometric spacing of decay rates is fixed entirely by the anomalous scaling exponent and remains insensitive to microscopic details of the short-distance regularization, which enter only through a single matching scale. This log-periodic spectrum is the direct spectral signature of a renormalization-group limit cycle, mirroring the structure familiar from Efimov physics and establishing the inverse-square interaction as a canonical generator of anomaly-driven dissipative hierarchies. In this sense, the emergent effective temperature is not a model-dependent construct but a universal kinematic scale arising from conformal symmetry breaking by irreversible boundary conditions.

\section{Analogy with black-hole quasi-normal modes}
The structure of the complex energy spectrum obtained in Sec.~III admits a natural comparison with black-hole quasi-normal modes (QNMs). In gravitational systems, QNMs arise as solutions of linearized field equations on a black-hole background subject to boundary conditions of purely ingoing waves at the event horizon and purely outgoing waves at spatial infinity. These conditions lead to a discrete set of complex frequencies whose imaginary parts govern the decay of perturbations and characterize the dissipative nature of black holes as open systems~\cite{Berti2009}.

A closely related structure appears in the present non-Hermitian Klein--Gordon model. The inverse-square interaction obtained in Eq.~(\ref{Sch_like}) plays a role analogous to the near-horizon effective potential in black-hole perturbation theory, where inverse-square terms arise from the centrifugal barrier and the near-horizon redshift in tortoise coordinates. In both cases, the local dynamics exhibits an emergent one-dimensional scale invariance that governs the short-distance behavior of the solutions.

In fact, the physical selection principle for the spectrum is the same in both systems. In black-hole physics, the causal structure of spacetime enforces a purely ingoing boundary condition at the horizon, reflecting the one-way nature of the event horizon. In the present model, the imaginary scalar potential renders the singular point $r=0$ a perfect absorber, and the purely outgoing boundary condition imposed in Sec.~III enforces directed flux into this absorbing center. In both cases, the boundary condition is dictated by physical irreversibility rather than by mathematical convenience.

As a result, the spectral problem in both settings admits a discrete set of complex frequencies that describe damped oscillations. In the present model, the spectrum takes the explicit log-periodic form
\begin{equation}
E_n = E_0 \exp\left(-\frac{\pi n}{\sigma_\ell}\right),
\end{equation}
with decay rates $\Gamma_n = -2\,\mathrm{Im}\,E_n$. While generic black-hole QNM spectra are not strictly log-periodic over the full frequency range, geometric spacing emerges in several analytically controlled limits, including near-extremal horizons and models with exact conformal symmetry in the near-horizon region~\cite{Hod1998}.

It is important to emphasize that the present construction does not describe gravitational dynamics or curved spacetime. The model is instead formulated entirely in flat spacetime, where horizon-like absorbing boundary conditions and the inverse-square conformal symmetry are realized within a minimal and exactly solvable effective framework. In this sense, the analogy with black-hole quasinormal modes is purely kinematical: the present analysis isolates the essential ingredients responsible for dissipative resonance spectra independently of their geometric realization in curved black-hole spacetimes.

From this viewpoint, the inverse-square anomaly, recast here as anomaly-driven dissipation, plays a role analogous to that of gravitational anomalies in near-horizon effective field theories of Hawking radiation. In both cases, a classically conserved symmetry becomes anomalous upon imposing irreversible boundary conditions, and the anomaly manifests itself as a universal decay or flux spectrum.

The present model thus provides a controlled analytic laboratory for studying how conformal symmetry, anomaly, and irreversible boundary conditions conspire to produce quantized dissipation, in close qualitative analogy with black-hole quasi-normal ringing.

\section{Conclusions}

In this work, we have investigated the Klein--Gordon equation in the presence of an imaginary inverse-square potential describing a perfectly absorbing center. We have shown that the singular inverse-square interaction retains the characteristic quantum anomaly of the Hermitian problem, but its physical manifestation is fundamentally altered by the absorbing boundary condition. Rather than producing a tower of bound states, the anomaly appears as a discrete hierarchy of decaying resonances characterized by a universal geometric scaling of the decay rates.

A central result of this work is the establishment of a conceptual bridge between the Hermitian and non-Hermitian descriptions of the inverse-square interaction. We have shown that the one-parameter family of self-adjoint extensions associated with the inverse-square quantum anomaly admits a natural analytic continuation to a family of non-Hermitian absorbing boundary conditions by allowing the boundary phase to become complex. In this continuation, the standing-wave boundary condition of the Hermitian problem is continuously deformed into a purely absorbing boundary condition, while the analytically continued spectrum acquires complex resonance energies whose imaginary parts determine the decay rates of the open system.

The derivation presented in Appendix A demonstrates that the geometric spectrum originates solely from the logarithmic short-distance structure of the inverse-square potential and the corresponding boundary condition at the singularity, independently of the absorbing dynamics. Appendix B then shows how this anomalous spectrum is reinterpreted in the non-Hermitian framework through analytic continuation, providing the physical realization of the same discrete scale invariance in terms of resonant decay rather than bound-state quantization. In this way, the inverse-square quantum anomaly is seen to survive the transition from a conservative to an open dissipative system without altering its underlying discrete scaling structure.

The principal observable prediction of the present theory is therefore not an absolute decay scale but the universal ratio between successive resonance widths,
\[
\Gamma_{n+1}=\Gamma_n\exp\!\left(-\frac{\pi}{\sigma_\ell}\right),
\]
which depends only on the strength of the inverse-square interaction and is independent of the overall normalization. This discrete scaling law constitutes the direct non-Hermitian analogue of the geometric spectrum first identified by Case for the Hermitian inverse-square Hamiltonian.

More generally, the present work shows that discrete scale invariance is not restricted to self-adjoint quantum systems but persists in open systems governed by non-Hermitian dynamics. This is achieved through an analytic continuation of the one-parameter family of self-adjoint boundary phases to a family of absorbing boundary conditions, preserving the underlying anomalous scaling while transforming the geometric tower of bound states into a hierarchy of dissipative resonances. The resulting framework provides a unified description connecting the inverse-square quantum anomaly, absorbing boundary conditions, and resonance decay, and may offer a useful perspective for the study of singular non-Hermitian interactions and other open wave systems exhibiting scale-anomalous behavior.

The present formulation also admits a natural physical interpretation.
Although the reduced field dynamics is non-unitary, it naturally admits a unitary dilation interpretation in which the apparent probability loss is reinterpreted as information transfer to an unobserved sector associated with the absorber. In this sense, the model provides a minimal analytic laboratory for studying anomaly-driven decay and horizon-like information flow in relativistic open quantum systems, isolating the kinematic ingredients responsible for dissipative resonance spectra without invoking gravitational dynamics.
Beyond its conceptual significance, the predicted log-periodic hierarchy of decay rates and the associated emergent effective temperature are, in principle, experimentally accessible through resonance spectroscopy and time-resolved decay measurements in synthetic platforms. Owing to its exact solvability, the present framework also provides a natural starting point for extensions to fermionic fields, gauge interactions, and curved spacetime backgrounds.

\appendix
\section{Self-Adjoint Extension and the Universal Geometric Spectrum}
\label{app:anomaly}

The purpose of this appendix is to derive the universal geometric spectrum associated with the supercritical inverse-square quantum anomaly, independently of the absorbing boundary condition introduced in the main text. To this end, we employ the self-adjoint extension framework pioneered by Case, which provides the standard Hermitian realization of the singular inverse-square Hamiltonian. Our goal is not to reproduce the complete matching calculation of Ref.~\cite{Case1950}, but rather to isolate the essential ingredients responsible for the emergence of the discrete geometric hierarchy: the logarithmic near-origin solutions, the non-essential self-adjointness of the Hamiltonian, and the introduction of a single ultraviolet scale through the self-adjoint extension. These ingredients are sufficient to establish the universal spectrum
\begin{equation}
E_n = E_0 \exp\!\left(-\frac{\pi n}{\sigma_\ell}\right),
\end{equation}
which depends only on the conformal anomaly. The complementary problem considered in the present work, namely the replacement of the self-adjoint boundary condition by a perfectly absorbing non-Hermitian boundary condition, is treated independently in \ref{app:absorber}.

\subsection{Exact Bessel Representation}
The exact solution of the radial equation is obtained by introducing the dimensionless variable
\begin{equation}
z=Er .
\end{equation}
Writing
\begin{equation}
u_\ell(r)=\sqrt{r}\,y(z),
\end{equation}
the radial equation becomes
\begin{equation}
y''+\frac1z y'+\left(1-\frac{\nu^2}{z^2}\right)y=0,
\end{equation}
where
\begin{equation}
\nu=i\sigma_\ell .
\end{equation}
The exact solution is therefore
\begin{equation}
u_\ell(r)=\sqrt r\left[A\,J_{i\sigma_\ell}(Er)+B\,J_{-i\sigma_\ell}(Er)\right].
\label{exact_bessel}
\end{equation}
Using the small-argument expansion
\begin{equation}
J_{\pm i\sigma_\ell}(Er)\sim\frac{(Er/2)^{\pm i\sigma_\ell}}{\Gamma(1\pm i\sigma_\ell)},\qquad Er\ll1 ,
\end{equation}
one recovers the near-origin form (\ref{power_solution}). The exact Bessel solution thus reproduces the logarithmic oscillations responsible for the anomaly and provides the global continuation of the ultraviolet solutions to finite radius.

\subsection{Near-Origin Behavior}
The radial equation \eqref{radial} may be written as
\begin{equation}
H_\ell u_\ell(z)=u_\ell(z),\qquad H_\ell=-\frac{d^2}{dz^2}-\frac{\alpha_\ell}{z^2},
\label{app_hamiltonian}
\end{equation}
where
\begin{equation}
\alpha_\ell=\gamma^2-\ell(\ell+1).
\end{equation}
Near the singular point $r=0$, the inverse-square interaction dominates over the finite spectral term $E^2$, and the radial equation reduces to
\begin{equation}
\frac{d^2u_\ell}{dz^2}+\frac{\alpha_\ell}{z^2}u_\ell\simeq 0 .
\label{uv_equation}
\end{equation}
Seeking a Frobenius solution
\begin{equation}
u_\ell(r)\sim z^s ,
\end{equation}
one obtains the indicial equation
\begin{equation}
s(s-1)+\alpha_\ell=0 ,
\label{indicial}
\end{equation}
where
\begin{equation}
\alpha_\ell=\gamma^2-\ell(\ell+1).
\end{equation}
The two exponents are
\begin{equation}
s_\pm=\frac12\pm\sqrt{\frac14-\alpha_\ell}.
\end{equation}
For
\begin{equation}
\alpha_\ell>\frac14 ,
\end{equation}
the square root becomes imaginary. Defining
\begin{equation}
\sigma_\ell=\sqrt{\alpha_\ell-\frac14},
\label{sigma_def}
\end{equation}
the general solution takes the form
\begin{equation}
u_\ell(z)=A\,z^{1/2+i\sigma_\ell}+B\,z^{1/2-i\sigma_\ell}.
\label{power_solution}
\end{equation}
Since the near origin solution should be matched with the exact solution \eqref{exact_bessel} we have $z=Er$ or \eqref{power_solution} equivalently is
\begin{equation}
u_\ell(r)=N\,r^{1/2}\cos\!\left(\sigma_\ell\ln \frac{r}{r_0}+B\right),
\label{log_solution}
\end{equation}
where $B=\left(\sigma_\ell\ln Er_0+\delta\right)$, $N$ is the normalization constant  and $\delta$ is a constant phase.
Equation~(\ref{log_solution}) exhibits logarithmic oscillations in the radial coordinate. The continuous scale symmetry of the classical inverse-square interaction is therefore replaced by a periodic structure in $\ln r$, signalling the onset of the inverse-square quantum anomaly. The real parameter $\delta$ is not determined by the differential equation itself and must be fixed by a short-distance boundary condition. This introduces a new physical scale into the problem and leads to dimensional transmutation and the emergence of a geometrically spaced spectrum.

\subsection{Self-Adjoint Extension and Quantization}
To determine the spectrum, consider two eigenfunctions for $E^2=\Lambda$,
\begin{equation}
H_\ell u_1=\Lambda_1 u_1,\qquad H_\ell u_2=\Lambda_2 u_2 .
\end{equation}
Self-adjointness requires
\begin{equation}
\langle u_2,H_\ell u_1\rangle- \langle H_\ell u_2,u_1\rangle= 0.
\end{equation}
After integrating by parts, this condition reduces to the vanishing of the boundary form
\begin{equation}
\left[u_2^*(r)\frac{du_1}{dr}-\frac{du_2^*}{dr}u_1(r)\right]_{r=0}=0 ,
\label{boundary_form}
\end{equation}
which leads to 
\begin{equation}
\sin\!(B_2-B_1=\sigma_\ell\ln\frac{E_2}{E_1})=0.
\end{equation}
Therefore $B_2-B_1=n\pi$ or all eigenfunctions belong to the same self-adjoint extension if
\begin{equation}
\sigma_\ell\ln(\frac{E_2}{E_1})=\pm n\pi,\qquad n \in \mathbb{Z}_+.
\label{boundary_self}
\end{equation}
Since the self-adjoint boundary condition is common to all eigenstates, the relation \eqref{boundary_self} holds for any pair of admissible eigenvalues $E_i$ and $E_j$.  Hence the spectrum is invariant under the discrete scaling transformation
\begin{equation}
E\rightarrow E\exp\left(\frac{\pi}{\sigma}\right),
\label{discrete_scaling}
\end{equation}
where choosing an arbitrary reference eigenvalue $E_0$ leads to the complete tower as
\begin{equation}
E_n = E_0 \exp\left(\pm \frac{\pi n}{\sigma_\ell}\right), \qquad n = 0, 1, 2, \ldots ,
\label{Lambda_spectrum}
\end{equation}
or equivalently
\begin{equation}
E_{n+1} = E_n \exp\left(\pm \frac{\pi}{\sigma_\ell}\right).
\end{equation}
The upper sign ($+$) corresponds to the fall-to-the-center branch where eigenvalues grow without bound, representing the instability that our non-Hermitian model converts into quantized dissipation. The lower sign ($-$) corresponds to the opposite direction of the geometric tower. Equation~(\ref{Lambda_spectrum}) is the characteristic bidirectional geometric spectrum of the inverse-square quantum anomaly. 

\section{Non-Hermitian Generalization of Case's Self-Adjoint Extension}
\label{app:absorber}

The geometric spectrum derived in \ref{app:anomaly} follows from the Hermitian self-adjoint extension of Case, characterized by a real boundary phase $\delta$. In the present work, the underlying Klein--Gordon equation is explicitly non-Hermitian due to the imaginary scalar potential. We now demonstrate that the perfectly absorbing boundary condition and the resulting resonance spectrum are not independent postulates, but rather the natural analytic continuation of Case's one-parameter family of self-adjoint extensions. 

\subsection{Complexification of the boundary phase and the perfect absorber}

In Case's Hermitian construction, the near-origin solution is a standing wave in the logarithmic coordinate [cf.\ Eq.~(\ref{log_solution})]:
\begin{equation}
u_\ell(r) = N\,r^{1/2}\cos\!\left(\sigma_\ell\ln Er + \delta\right),
\qquad \delta \in \mathbb{R}.
\end{equation}
Since the Klein--Gordon equation with the imaginary potential is non-Hermitian, there is no physical requirement that the extension parameter remain real. We promote $\delta$ to a complex value $\delta = \delta_R + i\delta_I$. 

Using the exponential representation of the cosine, the wavefunction decomposes into outgoing ($r^{1/2+i\sigma_\ell}$) and ingoing ($r^{1/2-i\sigma_\ell}$) components:
\begin{equation}
u_\ell(r) = \frac{N}{2}\,e^{-\delta_I}\,e^{i\delta_R}\,r^{1/2+i\sigma_\ell} \;+\; \frac{N}{2}\,e^{\delta_I}\,e^{-i\delta_R}\,r^{1/2-i\sigma_\ell}.
\end{equation}
As $\delta_I \to +\infty$, the coefficient of the ingoing wave diverges. However, this is merely an artifact of the overall normalization $N$. We therefore introduce a renormalized amplitude $\tilde{N} \equiv N e^{\delta_I}$, which remains finite. In terms of $\tilde{N}$, the wavefunction is:
\begin{equation}
u_\ell(r) = \frac{\tilde{N}}{2}\,e^{-2\delta_I}\,e^{i\delta_R}\,r^{1/2+i\sigma_\ell} \;+\; \frac{\tilde{N}}{2}\,e^{-i\delta_R}\,r^{1/2-i\sigma_\ell}.
\end{equation}
We identify the two components:
\begin{itemize}
\item \textbf{Outgoing wave} ($r^{1/2+i\sigma_\ell}$): amplitude $C_+ = \frac{\tilde{N}}{2}\,e^{-2\delta_I}\,e^{i\delta_R}$,
\item \textbf{Ingoing wave} ($r^{1/2-i\sigma_\ell}$): amplitude $C_- = \frac{\tilde{N}}{2}\,e^{-i\delta_R}$.
\end{itemize}

The ratio of outgoing to ingoing amplitudes is:
\begin{equation}
\frac{C_+}{C_-} = e^{2i\delta_R}\,e^{-2\delta_I}.
\label{ratio}
\end{equation}
The corresponding radial probability current, computed from $j_r = \sigma_\ell(|C_+|^2 - |C_-|^2)$, is:
\begin{equation}
\boxed{j_r = -\frac{|\tilde{N}|^2}{4}\,\sigma_\ell\,(1-e^{-4\delta_I}).}
\label{current}
\end{equation}
Equation~(\ref{current}) reveals that $\delta_I$ acts as a continuous dial controlling the absorption strength at the origin:

\begin{itemize}
\item \textbf{Hermitian limit} ($\delta_I = 0$): $j_r = 0$, $|C_+/C_-| = 1$. The wave is a pure standing wave, and probability is conserved. This recovers Case's self-adjoint extension exactly.

\item \textbf{Partial absorption} ($\delta_I > 0$): $j_r < 0$, $|C_+/C_-| = e^{-2\delta_I} < 1$. There is a net inward flux; the origin acts as a partial absorber with reflection coefficient $\mathcal{R} = e^{-4\delta_I}$.

\item \textbf{Perfect absorber} ($\delta_I \to +\infty$): $|C_+/C_-| \to 0$, $j_r < 0$. The outgoing component is completely suppressed, and the wavefunction reduces to:
\begin{equation}
u_\ell(r) \;\xrightarrow{\delta_I \to +\infty}\; \frac{N}{2}\,e^{\delta_I}\,e^{-i\delta_R}\,r^{1/2-i\sigma_\ell} \;\propto\; r^{1/2-i\sigma_\ell}.
\label{pure_ingoing}
\end{equation}
This is precisely the purely ingoing wave identified in the main text as the perfect absorber boundary condition.
\end{itemize}

Thus, the absorbing boundary condition is \emph{not} an independent physical postulate. It is the $\delta_I \to +\infty$ limit of the complexified Case extension. The non-Hermitian Klein--Gordon equation simply selects this limiting case as the physically relevant one, since the imaginary potential $i\gamma/r$ renders the origin a one-way sink.

\subsection{Analytic continuation of the asymptotic solution and the Jost function}

In Case's Hermitian bound-state problem, the exact radial solution that decays at infinity is given by the modified Bessel function of the second kind, $K_{i\sigma_\ell}(\kappa r)$, where $\kappa = \sqrt{-E^2} > 0$. 

When we analytically continue the energy to the complex plane to describe decaying resonances ($E \to E_R - i\Gamma/2$), the spectral parameter $\kappa$ becomes complex. The modified Bessel function $K_{i\sigma_\ell}(z)$ analytically continues to the Hankel function of the first kind:
\begin{equation}
K_{i\sigma_\ell}(-i z) \propto H^{(1)}_{i\sigma_\ell}(z).
\end{equation}
The Hankel function $H^{(1)}_{i\sigma_\ell}(Er)$ is the Jost solution, which is purely outgoing at spatial infinity. 

Therefore, the analytic continuation of the global bound-state solution transforms the exponentially decaying asymptotic behavior into the outgoing Jost solution. At the same time, the continued boundary phase continuously deforms the near-origin standing wave into the purely ingoing logarithmic solution corresponding to a perfect absorber. In this way, the Hermitian bound-state problem is continuously connected to the non-Hermitian resonance problem.

\subsection{Geometric spectrum of decay rates}

The bound-state eigenvalues in Case's framework form a geometric tower $E_n = E_0 \exp(\pm\pi n/\sigma_\ell)$. 
In fact, Eq.~(\ref{Lambda_spectrum}) follows entirely from the inverse-square singularity and the requirement of self-adjointness. No absorbing boundary condition is imposed. Here, we show that for the absorbing boundary condition by selecting  the exponentially decaying  branch ($\mathrm{Im}\,E<0$) and maps it into the complex energy plane as a hierarchy of decaying resonances, preserving the runaway character of the fall-to-the-center instability in the form of discrete dissipation.  To this end, by analytic continuation to the complex energy plane, the complex resonance energies are $E_n = \omega_n - i\Gamma_n/2$, where the decay widths can be obtained as
\begin{equation}
 \Gamma_n = -2\,\mathrm{Im}\,E_n=-2\exp(-\pi n/\sigma_\ell)\,\mathrm{Im}\,E_0=\Gamma_0\, \exp(-\pi n/\sigma_\ell), 
\end{equation}
where $E_0$ is a complex reference scale.
 Because the geometric ratio is multiplicative, the decay rates exhibit the exact same universal geometric spacing:
\begin{equation}
\frac{\Gamma_{n+1}}{\Gamma_n} = \exp\left(-\frac{\pi}{\sigma_\ell}\right).
\end{equation}
The dissipation thus proceeds in a quantized geometric progression, directly inherited from the Hermitian fall-to-the-center instability via analytic continuation.

In conclusion, the absorber boundary condition is the natural, physically motivated endpoint of the complexified self-adjoint extension family. The inverse-square anomaly, when embedded into a non-Hermitian relativistic framework, inevitably produces quantized dissipation with universal geometric scaling which reflects the limit-cycle structure of the renormalized theory.

\subsection{Robustness against subleading perturbations}

The geometric scaling ratio between consecutive levels is:
\begin{equation}
\frac{E_{n+1}}{E_n} = \exp\left(-\frac{\pi}{\sigma_\ell}\right).
\end{equation}
Crucially, the exponent $\sigma_\ell$ arises \emph{solely} from the leading $1/r^2$ singularity in Eq.~(\ref{uv_equation}). If a generic perturbation $\delta V(r)$ satisfying $\delta V(r) \ll \gamma/r$ as $r \to 0$ is introduced, it modifies the short-distance matching phase $\theta_{\text{UV}}$, thereby shifting the reference scale $E_0$. However, because the perturbation is subleading, it does not alter the order of the Bessel functions, and thus \emph{cannot change} $\sigma_\ell$.  Theoretically, the quantum anomaly and the resulting log-periodic spectrum are governed entirely by the short-distance (UV) behavior of the effective radial potential. Consider a generic perturbation $\delta V(r)$ arising from self-energy corrections or environmental coupling, such that the effective mass term becomes $m + V(r) = i\gamma/r + \delta V(r)$. Squaring this term, as it appears in the Klein--Gordon equation, yields:
\begin{equation}
\big(m + V(r)\big)^2 = -\frac{\gamma^2}{r^2} + \frac{2i\gamma \delta V(r)}{r} + \big(\delta V(r)\big)^2.
\end{equation}
For the $1/r^2$ singularity to remain the dominant UV feature, the perturbation must satisfy the condition $\delta V(r) \ll \gamma/r$ as $r \to 0$. This ensures that both the linear cross-term ($2i\gamma \delta V(r)/r$) and the quadratic term ($(\delta V(r))^2$) are strictly subleading to the dominant $-\gamma^2/r^2$ term. 

We can classify realistic perturbations into three regimes based on this condition:
\begin{enumerate}
    \item $\delta V(r) = \text{constant}$: This trivially satisfies $\delta V(r) \ll \gamma/r$ as $r \to 0$, merely shifting the effective energy offset.
    \item $\delta V(r)$ is $r$-dependent but less singular than $1/r$ (e.g., regular at the origin including positive powers of $r$ ). This also strictly satisfies the subleading condition.
    \item $\delta V(r) \propto 1/r$: In this borderline case, the perturbation does not introduce a new singularity; it merely renormalizes the effective coupling constant $\gamma \to \gamma_{\text{eff}}$. The $1/r^2$ structure is perfectly preserved, and the log-periodic scaling remains intact (with a slightly shifted, but still constant, exponent $\sigma_\ell$).
\end{enumerate}
Cases more singular than $1/r$ (e.g., $\sim 1/r^2$ or stronger) are unphysical for standard scalar interactions. For all physically realistic cases (1--3), standard theorems on singular differential operators (Weyl’s limit point/limit circle classification \cite{ReedSimon}) guarantee that the deficiency indices of the radial Hamiltonian are unchanged. 

This UV dominance has a profound implication for the stability of the spectrum. Any modification to the potential that satisfies $\delta V(r) \ll \gamma/r$ will alter the short-distance phase, thereby shifting the overall reference energy scale $E_0$. However, because the leading $1/r^2$ singularity remains strictly dominant (or merely renormalized), the functional form of the spectrum is preserved. Consequently, the geometric scaling ratio $E_{n+1}/E_n = \exp(-\pi/\sigma_\ell)$ is a universal, robust feature of the theory, entirely insensitive to the specific form of the potential at large scales or to mild short-distance corrections. This confirms that the discrete scale invariance is an intrinsic property of the UV singularity, fundamentally decoupled from the IR details of the system. This invariance follows from Weyl's theory of singular differential operators: subleading perturbations preserve the deficiency indices and the leading indicial exponent, affecting only the boundary phase that determines the reference scale $E_0$.

\end{document}